# Towards Automation of Creativity: A Machine Intelligence Approach

Subodh Deolekar[1,*], Siby Abraham[2]

[1]Department of Computer Science, University of Mumbai, Mumbai 98, India. subodhdeolekar@gmail.com
[2]Department of Maths & Stats, G. N. Khalsa College, University of Mumbai, India. sibyam@gmail.com

**Abstract:** This paper demonstrates emergence of computational creativity in the field of music. Different aspects of creativity such as producer, process, product and press are studied and formulated. Different notions of computational creativity such as novelty, quality and typicality of compositions as products are studied and evaluated. We formulate an algorithmic perception on human creativity and propose a prototype that is capable of demonstrating human-level creativity. We then validate the proposed prototype by applying various creativity benchmarks with the results obtained and compare the proposed prototype with the other existing computational creative systems.

## I. INTRODUCTION

Computational creativity is the modeling or replicating human creativity computationally. Traditionally computational creativity has focused more on creative systems' products or processes, though this focus has widened recently. Research on creativity offers four Ps of creativity (Rhodes, 1961; MacKinnon, 1970; Jordanous, 2016).

These four P's are:

1. Person/Producer: a creative agent

2. Process: an activity done by the creative agent

3. Product: the product of the creative process

4. Press/Environment: the overall environment of creativity

The proposed methodology addresses all the four P's of creativity unlike most of recent works, which focus on these individually (Saunders, 2012; Gervas & Leon, 2014; Misztal & Indurkhya, 2014; Sosa & Gero, 2015; Besold & Plaza, 2015; Harmon, 2015). Figure 1 gives a simplified view of proposed computational creative system in the context of four P's of creativity.

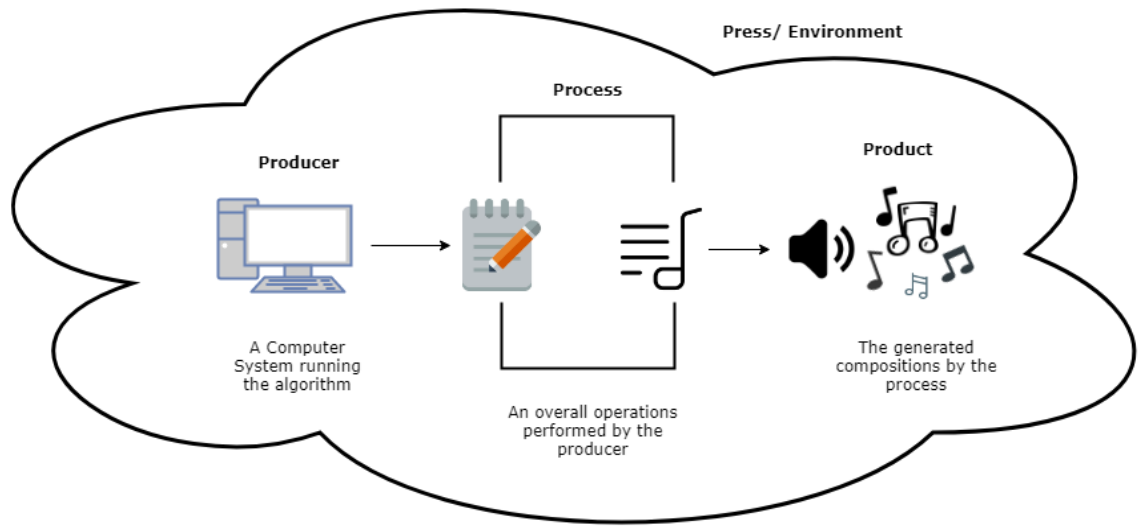

**Fig. 1.** Simplified overview in terms of four P's for proposed computational creativity model

**1. Person / Producer**

The creative producer in our system is a computer system which implements the algorithm for creativity in music composition. Since domain knowledge is important when a producer is given a task to anticipate necessary skills of creativity, we provide producer with a music composition and its type, based on which the producer is able to generate the desired rule set.

**2. Process**

The creative process involved here comprises of generating quality compositions using techniques such as Markov chain and memetic algorithm. The process includes other subtasks such as extraction of rules set and selection of best compositions based on fitness value.

**3. Product**

The proposed system's creative product is improvised variations of the input composition. The variations are novel from each other in make but carry equivalent quality. The generated creative products are then evaluated using Graeme Ritchie's empirical criteria (Ritchie, 2001, 2007), which exclusively evaluates the products for proposed computational creative system.

**4. Press / Environment**

The notion of environment here is to provide interactions between producer and process to have creative product. In the proposed system the press is computational environment which provide controllable settings to have emergence of creativity. The environment influences the individual compositions with the help of creative process to create improvised variations which are creative products.

**II. PROPOSED METHODOLOGY**

To illustrate the approach of proposed system, we consider the rational starting point for an analysis of tabla music in a category known as kāyadā composition which, musicians believe, constitutes the very basis of improvisation in tabla playing (Courtney, 1994). The system takes input as kāyadā theme and generate variations. The generated variations are unique in nature where many of the variations are never seen before by the experts in the field. The system ranks the new variations according to their fitness value and selects the variations which are best among the population. Figure 2 demonstrates the overall process of providing theme composition to the system, generating different variations and evaluation of the same.

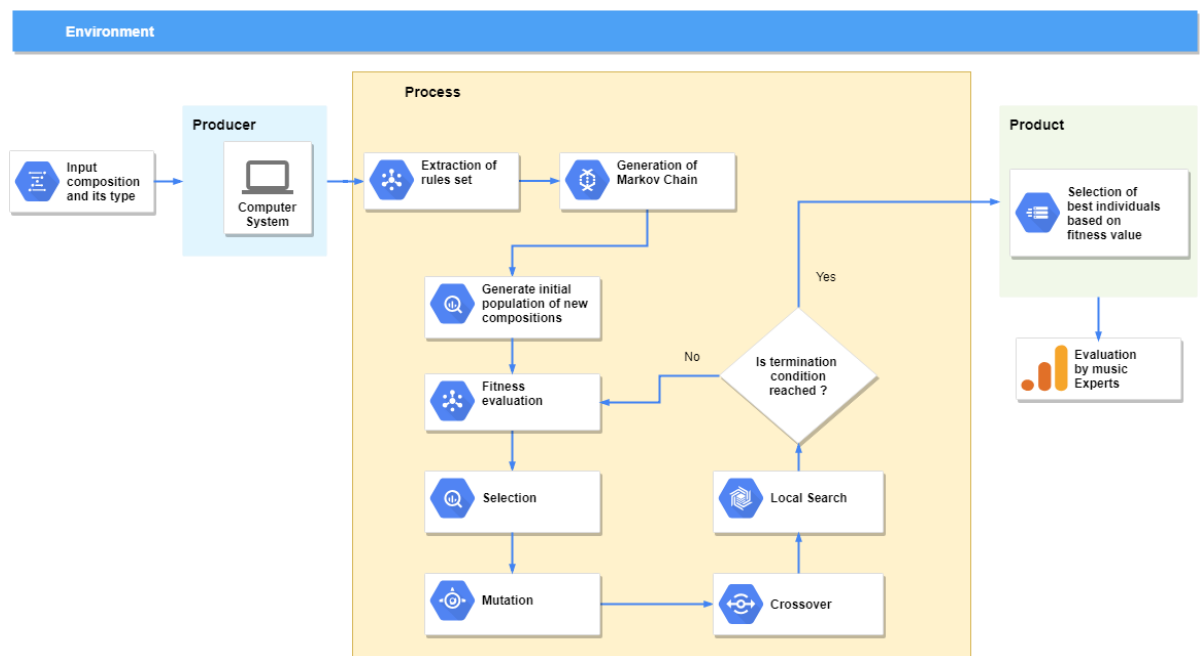

**Fig. 2.** System workflow

**Input composition**

The system takes an input composition and its type in textual format. The methodology works with any composition once its type is known. Here, we provide kāyadā theme composition as an input composition to the system with its type as cyclic (Stewart, 1974).

**Extraction of rules set**

We derive the rules set from the input composition. Based on the rules set the system tries to generate new variations of the theme composition. Every kāyadā composition has a specific phrase associated with its theme, which expresses the beauty of the composition; we call this phrase as highlighted phrase of that composition.

The theme of kāyadā which is mentioned in figure 3 has highlighted phrase as **'DhaTiT'**. The system extracts the highlighted phrase of the kāyadā composition. Since kāyadā is specified by its structure we can go over the rules and form different variations of the kāyadā theme.

| **DhaTi** | TDha | TiT | DhaDha | TiT | DhaGa | TinNa | KiNa |
|---|---|---|---|---|---|---|---|
| TaTi | TTa | TiT | TaTa | TiT | DhaGa | DhinNa | GiNa |

**Fig. 3.** Kāyadā theme

The very first rule of a kāyadā is that the bōls which are there in the theme of a kāyadā must be maintained in further variations of the same. If we look at the example specified in figures 4 and 5, we clearly see that the bōls which are present in the theme, only those bōls are present in the variations.

| DhaTi | TDha | DhaTi | TDha | TiT | DhaGa | TinNa | KiNa |
|---|---|---|---|---|---|---|---|
| TaTi | TTa | TaTi | TDha | TiT | DhaGa | DhiNa | GiNa |

**Fig. 4.** Variation 1

| DhaTi | TDha | TiT | DhaDha | TiT | TiT | TinNa | KiNa |
| TaTi | TTa | TiT | DhaDha | TiT | TiT | DhiNa | GiNa |

**Fig. 5.** Variation 2

**Markov Chain**

The methodology proposes the use of Markov Chain as an effective tool for creating new valid compositions and the idea of stochastic matrices work by analyzing the chance of any given bōl (to put the example in a musical context) going directly to any other bōl. We calculate this by checking a particular bōl every time it occurs in a composition and giving weights to those bōls, that follow it into a stochastic matrix. To do this, for every bōl value we build a stochastic matrix which resembles the composition that it is based on.

**Creation of Markov Matrix**

We consider a kāyadā theme composition as shown in Figure 3. We identify the unique bōls in the first half of the theme composition (Only first half i.e. bharī, which comprises of open sounds, is enough to generate the new composition as the other half is nothing but the khālī, comprise of closed sounds, part of the first one) and we form a two-dimensional matrix. The matrix shows the unique bōls which are there in the theme composition arranged in rows and columns. The left most column of the matrix, which are in bold, are possible 'seeds' and top most row is known as 'transitions' or 'output'. We fill the matrix by counting the frequency of all bōls in the left column, which are followed by a bōl in the top column.

**Table 1.** Markov matrix

|     | Dha | Ti | T | Ga | Tin | Na | Ki |
| --- | --- | --- | --- | --- | --- | --- | --- |
| **Dha** | 1 | 3 | 0 | 1 | 0 | 0 | 0 |
| **Ti**  | 0 | 0 | 3 | 0 | 0 | 0 | 0 |
| **T**   | 3 | 0 | 0 | 0 | 0 | 0 | 0 |
| **Ga**  | 0 | 0 | 0 | 0 | 1 | 0 | 0 |
| **Tin** | 0 | 0 | 0 | 0 | 0 | 1 | 0 |
| **Na**  | 1 | 0 | 0 | 0 | 0 | 0 | 1 |
| **Ki**  | 0 | 0 | 0 | 0 | 0 | 1 | 0 |

We generate a proportional frequency table, as shown in the table 2:

**Table 2.** Relative frequency table generated from table 1.

|     | Dha | Ti  | T | Ga  | Tin | Na | Ki  |
|-----|-----|-----|---|-----|-----|----|-----|
| Dha | 0.2 | 0.6 | 0 | 0.2 | 0   | 0  | 0   |
| Ti  | 0   | 0   | 1 | 0   | 0   | 0  | 0   |
| T   | 1   | 0   | 0 | 0   | 0   | 0  | 0   |
| Ga  | 0   | 0   | 0 | 0   | 1   | 0  | 0   |
| Tin | 0   | 0   | 0 | 0   | 0   | 1  | 0   |
| Na  | 0.5 | 0   | 0 | 0   | 0   | 0  | 0.5 |
| Ki  | 0   | 0   | 0 | 0   | 0   | 1  | 0   |

**Formation of Markov Chain**

**Identify Seeds**

Another important rule of a kāyadā composition is that the starting bōl of a variation should be a whole word. We call these words as 'seeds'.

```
DhaTi   TDha   DhaTi   TDha   TiT   DhaGa   TinNa   KiNa
TaTi    TTa    TaTi    TDha   TiT   DhaGa   DhiNa   GiNa
```

**Fig. 6.** Variation 3

In the example which is shown in Figure 6 the whole words are 'Dha' and 'Ti' which play important roles in the given theme. So, we identify the potential seeds of the variations and begin the new variations with one of these seeds (Figure 7).

```
DhaDha  TiT   DhaTi   TDha   TiT   DhaGa   TinNa   KiNa
TaTa    TiT   TaTi    TDha   TiT   DhaGa   DhiNa   GiNa
```

**Fig. 7.** Composition starting with seed as 'Dha'

**Supply Seed**

A Markov matrix provides the far more useful ability to find an output based on an input. The input is known as the seed. For example, consider the first row of the relative frequency table

2. For seed 'Dha', there is 20% likelihood that the output will be 'Dha', 20% likelihood that the output will be 'Ga' and 60% likelihood that the output will be 'Ti'. Following this one step further we can use the output of this as a seed of the next choice.

The choice of output is based on a random number being smaller than the sum of weighing's going left to right across the row. We get the Cumulative relatives from the above table as:

**Table 3.** Cumulative relatives generated from Table 2.

|     | Dha | Ti  | T | Ga  | Tin | Na | Ki  |
| --- | --- | --- | --- | --- | --- | --- | --- |
| Dha | 0.2 | 0.8 | 0 | 1.0 | 0   | 0  | 0   |
| Ti  | 0   | 0   | 1 | 0   | 0   | 0  | 0   |
| T   | 1   | 0   | 0 | 0   | 0   | 0  | 0   |
| Ga  | 0   | 0   | 0 | 0   | 1   | 0  | 0   |
| Tin | 0   | 0   | 0 | 0   | 0   | 1  | 0   |
| Na  | 0.5 | 0   | 0 | 0   | 0   | 0  | 1.0 |
| Ki  | 0   | 0   | 0 | 0   | 0   | 1  | 0   |

**Table 4.** Seed and its corresponding output value depending upon random number.

| Seed | Random Number | Output |
| --- | --- | --- |
| Dha | 0.3 | Ti |
| Ti  | 0.8 | T |
| T   | 0.9 | Dha |
| Dha | 0.3 | Ti |
| Ti  | 0.7 | T |
| T   | 0.2 | Dha |
| Dha | 0.6 | Ti |
| Ti  | 0.2 | T |
| T   | 0.9 | Dha |
| Dha | 0.1 | Dha |
| Dha | 0.9 | Ga |
| Ga  | 0.2 | Tin |
| Tin | 0.8 | Na |
| Na  | 1   | Ki |
| Ki  | 0.7 | Na |

We consider Markov chain as a Finite state machine in which the transitions are probability driven. Figure 8 demonstrates the composition representation in the form of finite state machine. The states are represented by bōls in a composition and transitions are by random number generated by the system.

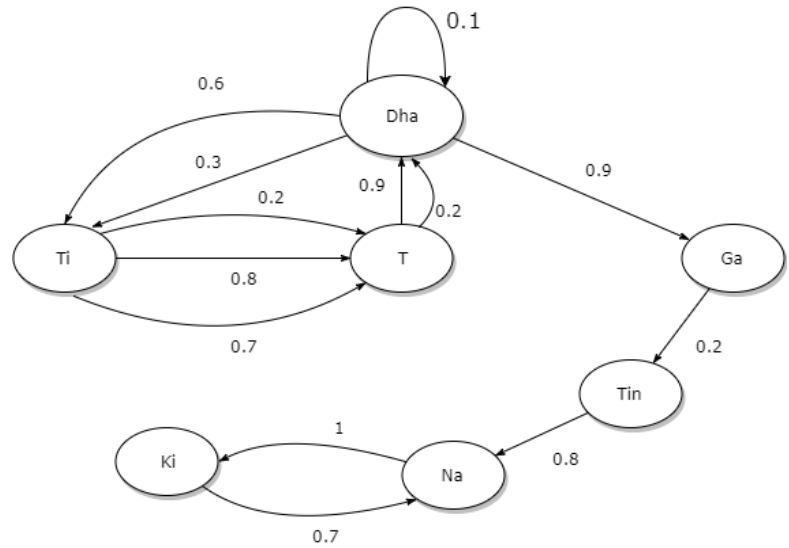

**Fig. 8.** Markov chain as a Finite state machine

The final sequence of generated set of symbols is considered as a composition and it looks like:

**'Dha Ti T Dha Ti T Dha Ti T Dha Dha Ga Tin Na Ki Na'**

**Memetic Algorithm**

We consider major four components of a memetic algorithm such as: initial population, a local search, crossover and mutation operation and regeneration of population.

**Initial Population**

The initialization process of memetic algorithm generates the 'N' members of the population which are based on the given theme of a kāyadā composition. Compositions which are generated using Markov chain are considered for the initial population. Table 5 shows the initial population generated for memetic algorithm.

**Table. 5.** Initial population generated from a kāyadā theme.

| Dha | Dha | Dha | Dha | Dha | Ga | Tin | Na | Ki | Na | Dha | Ga | Tin | Na | Ki | Na |
|---|---|---|---|---|---|---|---|---|---|---|---|---|---|---|---|
| Dha | Ti | T | Dha | Ti | T | Dha | Ti | T | Dha | Ti | T | Dha | Ga | Tin | Na |
| Dha | Dha | Ti | T | Dha | Ga | Dha | Ti | T | Dha | Ti | T | Dha | Ga | Tin | Na |
| Dha | Ti | T | Dha | Ti | T | Ti | T | Dha | Ga | Tin | Na | Ki | Na | Ki | Na |
| Dha | Ga | Tin | Na | Ti | T | Ti | T | Dha | Ti | T | Dha | Ti | T | Dha | Ti |

| Dha | Ti | T | Dha | Ti | T | Dha | Dha | Ti | T | Dha | Ti | T | Dha | Ga | Tin |
|-----|-----|-----|-----|-----|-----|-----|-----|-----|-----|-----|-----|-----|-----|-----|-----|
| Dha | Ti | T | Dha | Ti | T | Dha | Ti | T | Dha | Ga | Tin | Na | Dha | Dha | Ti |
| Dha | Ti | T | Dha | Ti | T | Dha | Dha | Ti | T | Dha | Ti | T | Dha | Ti | T |
| Dha | Dha | Ti | T | Dha | Ti | T | Dha | Ti | T | Dha | Ti | T | Dha | Dha | Dha |
| Dha | Dha | Ti | T | Dha | Ti | T | Dha | Ti | T | Dha | Ti | T | Dha | Ti | T |

**Fitness function**

We tried to formulate a fitness function which measures the aesthetic worth of a melody of a kāyadā composition and helps for evaluation to get a good candidate solution. We define a fitness function based on various rules specified for kāyadā composition.

**Identify the full verb**

As per the grammar of kāyadā composition, it should end with a full verb. We identify the possible full verb of a composition from a given theme. The kāyadā theme which is mentioned in figure 9 has verbs 'TinNaKiNa' and 'DhiNaGiNa'.

**Identify the half verb**

Many times, it is not always possible to have a full verb in the composition but a part of a full verb may be present. The compositions formed may not be a great composition but might be an improvised variation. We identify these possible half verbs. The kāyadā theme presented in figure 9 has half verbs as 'KiNa' and 'GiNa'.

**Identify the repetition of verb**

There must be a single verb in the kāyadā composition which comes at the end of the composition. Apart from the end location if a verb appears in middle of kāyadā composition then that composition is considered as invalid composition. The composition which is depicted in figure 9 is not a valid composition since it has repetition of a verb.

| DhaDha | TiT | DhaTi | TDha | TiT | DhaGa | TinNa | KiNa |
|--------|-----|-------|------|-----|-------|-------|------|
| TaTa | **TiNa** | **KiNa** | TDha | TiT | DhaGa | DhiNa | GiNa |

**Fig. 9.** Variation with repetition of verb

**Table. 6.** Rule set for kāyadā composition and its associated weight values

| Weight No. (Wi) | Rule | Value |
|---|---|---|
| 1 | Check for the Seed. Seed must be a whole word in kāyadā composition | 10 |
| 2 | End of a new composition with a full verb | 10 |
| 3 | End of a new composition with a half verb | 5 |
| 4 | There should not be more than one verb in the new composition | -2 for each repetition |

We mention the rules of kāyadā composition in table 6 which are checked and associated weights for those rules are applied for that composition. We take the sum of weights for which the kāyadā composition satisfies the rules as:

$$\theta_i = \sum_{j=1}^{4} W_j \qquad (1)$$

We call this as theta value, $\theta_i$ where i=1, 2, …, N. 'N' is a total number of compositions. The theta value for the theme composition is calculated as $\theta_0$ which is used as a base value. We use the fitness function for $i^{th}$ composition as shown in equation 2.

$$f_i = |\theta_0 - \theta_i| \qquad (2)$$

The equation 2 depicts the distance between the theme composition and the new generated improvised composition.

```
DhaTi  TDha  DhaTi  TDha  TiT  DhaGa  TinNa  KiNa
TaTi   TTa   TaTi   TDha  TiT  DhaGa  DhiNa  GiNa
```

**Fig. 10.** Variation of kāyadā composition

We calculate the values for the weights for a composition represented in figure 10. Weight w1 = 10 is assigned since the composition starts with 'Dha', w2 = 10 is assigned since it ends with the full verb 'Tin Na Ki Na'.

Since the full verb was present in the composition, the value for w3 is assigned as zero and lastly w4 is assigned as zero, as there is no repetition of the verb in the given composition. The total value of $\theta_1$ is evaluated as 20. The base value for theme composition is calculated with the same rules set and it comes around to be 20. As described in equation 2 the fitness value for the new composition is calculated as zero. The fitness function, proposed is a minimization function as we want to obtain a composition with minimum fitness value as the optimum solution.

**Local Search**

The purpose of local search is to enhance the quality of the solution by causing improvements in the individual solutions. The localized search for MA traverses through each candidate solution in the population, and improves the solution by selecting those which satisfies the different rules which are specified in the theme of kāyadā composition.

**Pseudo code for the local search**

**Step 1:** Identify the occurrence of full verb which is of length 4, in the composition except from its defined location (i.e. end of the composition)

    1.1 Replace the repeated verb with the highlighted phrase from the theme of length 4.
    1.2 Find the fitness of newly generated composition.

**Step 2:** Identify the occurrence of half verb (length 2) in the composition.

    2.1 Replace the half verb with the highlighted phrase from the theme of length 2.
    2.2 Find the fitness of the newly generated composition.

**Step 3:** Identify the occurrence of the length 4 phrase which contains the repetition of same bōl (e.g. 'DhaDhaDhaDha')

    3.1 Replace the phrase with the highlighted phrase from the theme of length 4.
    3.2 Find the fitness of newly generated composition.

The local search proposed in the methodology is illustrated below with an example:

1. If the full verb (of length 4) is found in the middle of the composition then replace those bōls with possible bōls like 'DhaTi T Dha', 'Ti T DhaDha', 'Ti T Dha Ga' and 'DhaDhaTi T' which are the highlighted phrases from derived from a given theme.

2. If the half verb (of length 2) is found in the middle of the composition then replace those bōls with the possible bōls like 'Ti T', 'DhaDha'.

3. If 'Dha' bōl is coming continuously 4 times, replace those bōls with the possible bōls like 'DhaDhaTi T', 'DhaTi T Dha', 'Ti T DhaDha', 'DhaTi T Dha'.

**Mutation**

We apply the multipoint mutation in MA in order to obtain better compositions. In the process of mutation, we identify random offspring for the mutation operation from the population. Mutation points are selected and the bōls within the mutation points are replaced with complementary bōls. The complimentary bōls for each bol is given table 7.

**Table. 7.** Bōls which are used as a complementary bōls in the khālī part

| Sr. No. | Bōl | Complementary bōl |
|---|---|---|
| 1 | Dha | Ta |
| 2 | Dhi | Tin |
| 3 | Gi | Ki |

The example described in figure 11 describes the mutation operation.

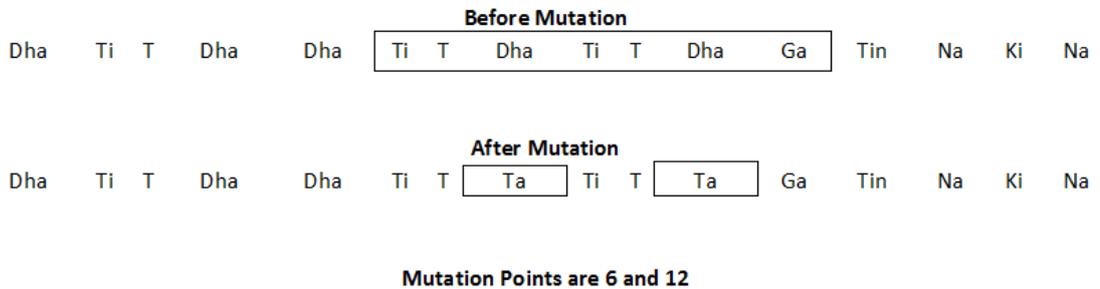

**Fig. 11.** Mutation operation

The aim of mutation process is to add diversity and to prevent the algorithm from poor local minima. After the fitness calculation we replace the worst solution in the population with the new offspring. The probability of a mutation operation selection is based on purely random bases which increases the search in other regions of the population.

**Crossover**

For crossover operation we randomly pair members (i.e. parents) from the mating pool and apply single point crossover operator to produce two offsprings from each pair. Newly created offspring are added to the population.

We randomly select the crossover point between 1 to $L$-1 where $L$ is the length of the chromosome. All the bōls between the crossover points are swapped and two new offsprings are produced. The crossover operation is depicted in the figure 12.

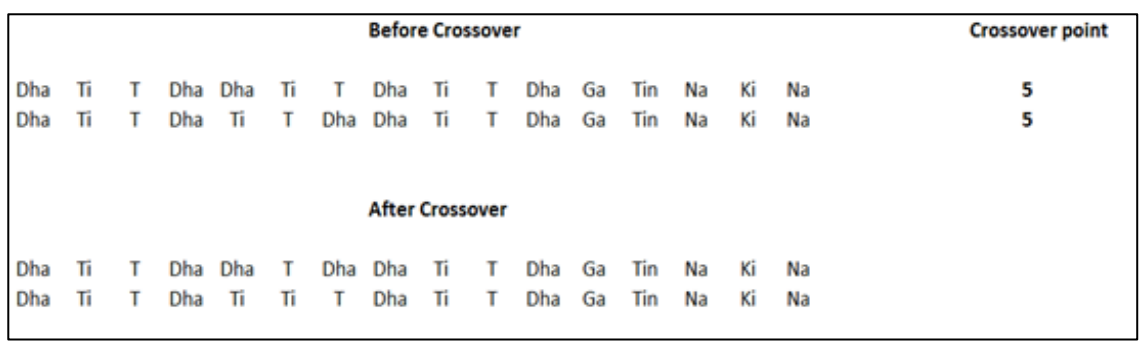

**Fig. 12.** Crossover operation

**Termination condition**

The algorithm stops when a termination condition is satisfied. This condition guarantees that the population fitness has reached the goal fitness value or that the number of iterations carried out so far has reached a predetermined limit.

**III. Results and discussion**

**Music Information Retrieval as creativity measure**

We wanted to check the contribution of features of audio signals of computer-generated variations for computational creativity. We considered flux, MFCC, brightness and novelty as a representative feature for the same. We extract envelop signal for the theme composition and its three variations which are of good quality. The figure 13 shows the envelope of theme composition and figures 14, 15 and 16 show the envelope of different variations.

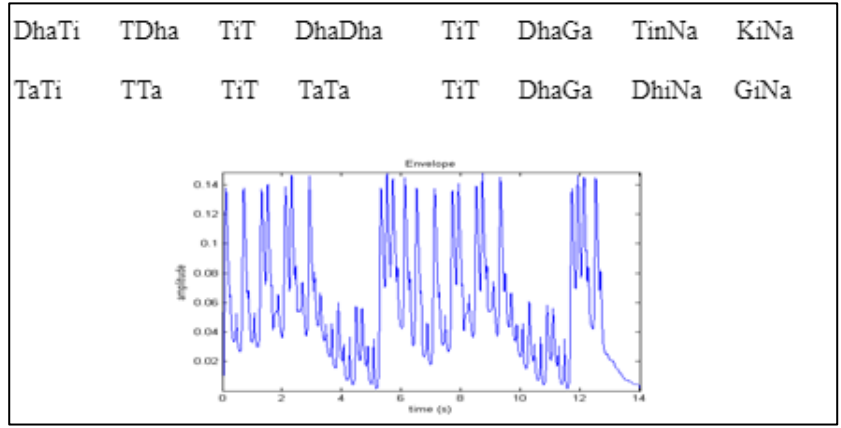

**Fig. 13.** Theme composition and its envelope

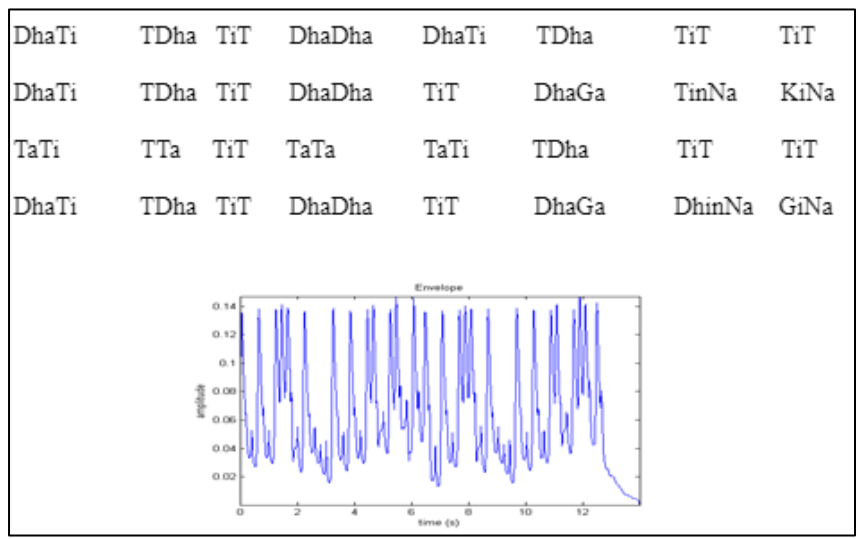

**Fig. 14.** Variation 1 and its envelope

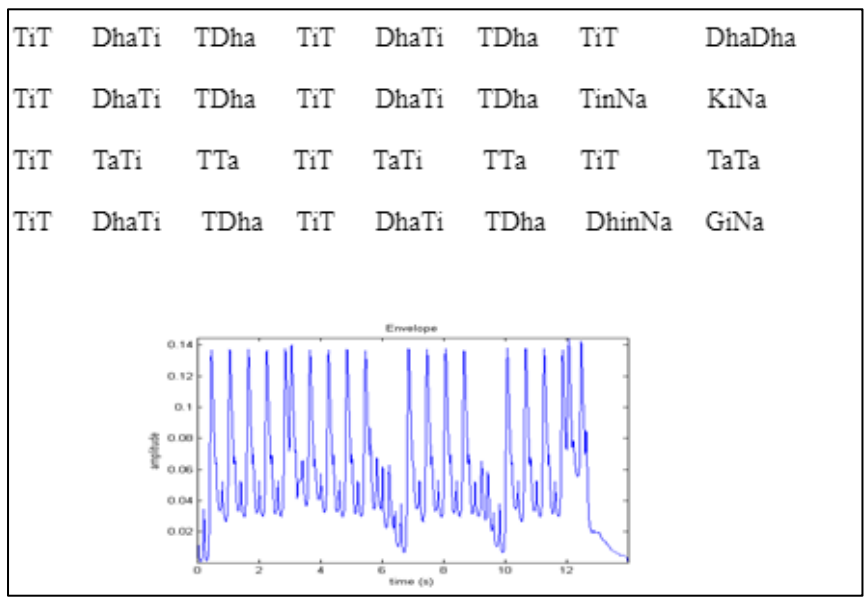

123

**Fig. 15.** Variation 2 and its envelope

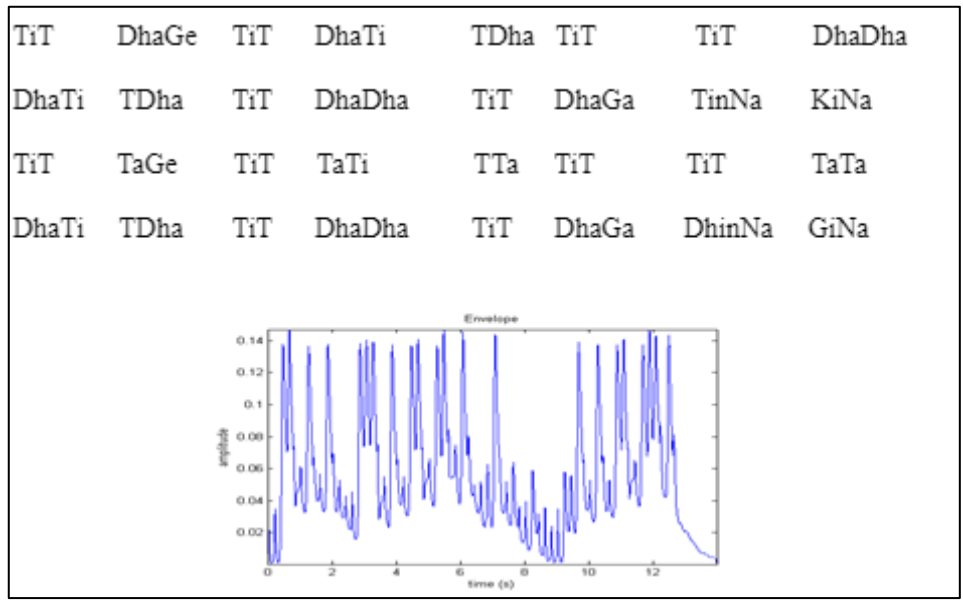

**Fig. 16.** Variation 3 and its envelope

These variations have different envelop shape as compared to the theme composition. But the quality of the variations is almost similar. We tried to extract the above-mentioned features of these compositions.

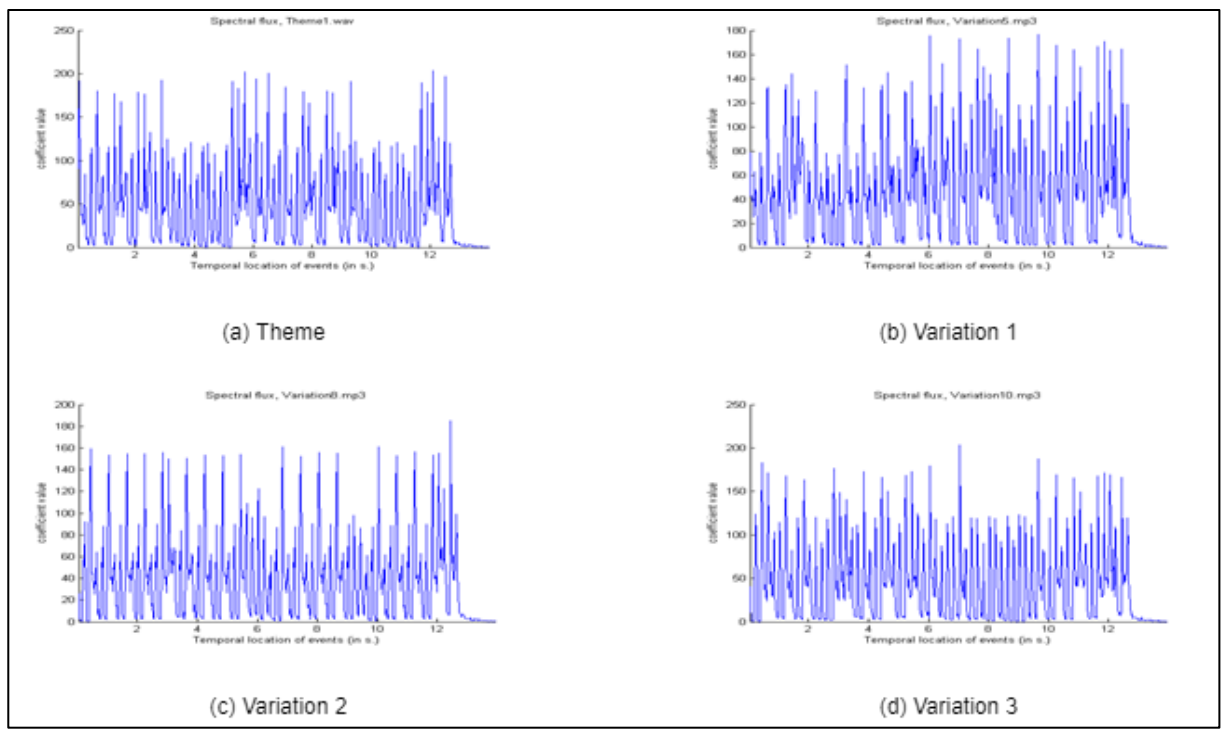

**Fig. 17.** Spectral Flux of a theme composition and its variations

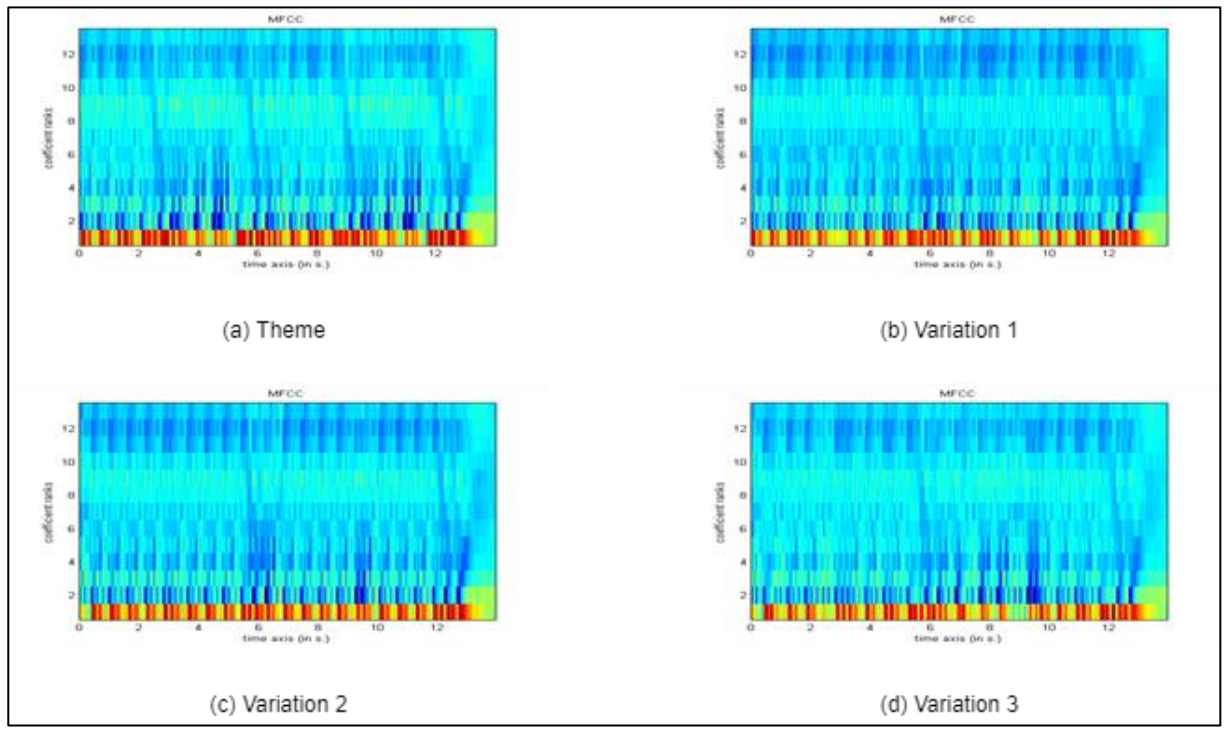

**Fig. 18.** MFCC of a theme composition and its variations

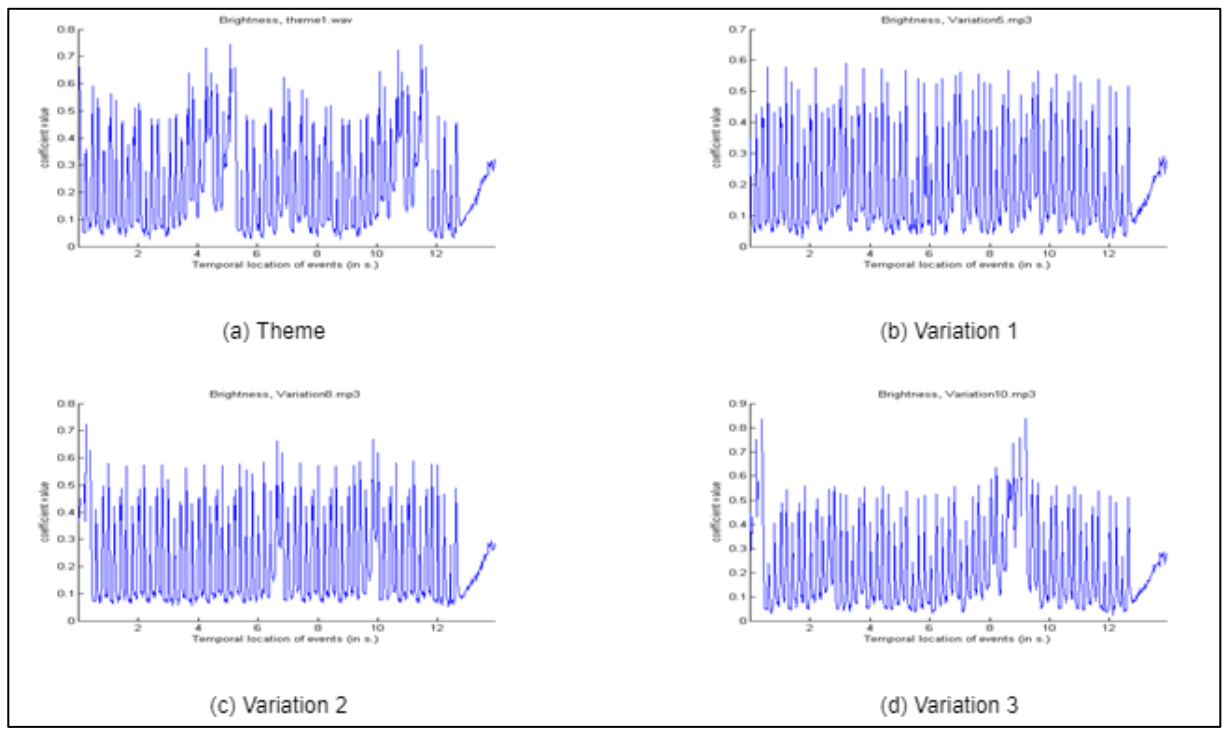

**Fig. 19.** Brightness of a theme composition and its variations

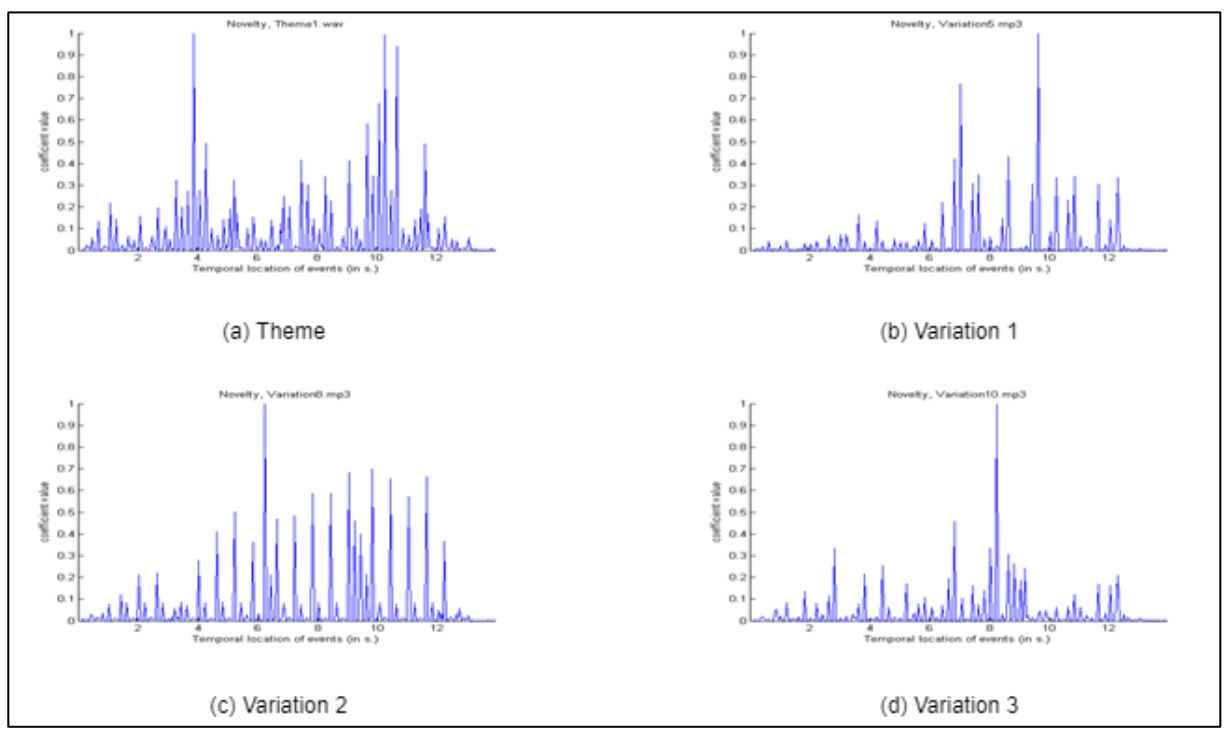

**Fig. 20.** Novelty of a theme composition and its variations

In figures 17 to 20, (a) part represents the various features extracted for the theme composition whereas, (b), (c) and (d) parts represent feature values obtained for variations 1, 2 and 3 respectively. By comparing these figures from 17 to 20 we could be able to say that (b), (c) and (d) are not only different from one another but are distinct from the original theme, therefore we could be able to get entirely different parameter values for the flux, MFCC, brightness and novelty.

Inherently they could be able to come with the different variations of the original theme but still they are the valid compositions. We can clearly see that a process could be able to generate the creative compositions by changing different features of theme.

The feature values on isolated scales are different but collectively they belong to the same class. Effectively the methodology could be able to try the different possible values for each of the features.

**Combinational, Exploratory and Transformational creativity**

In order to check creativity model proposed by (Boden, 1999), we investigate the proposed system with all three types of creativity. The first one is combinational creativity which involves fresh combination of familiar ideas. In our example, the kāyadā theme is given as the input to the system. Based on kāyadā theme the system generates new combinations of compositions which are called as palaṭās of that kāyadā.

Second one is, exploratory creativity which involves the generation of fresh ideas by the searching of structured conceptual spaces. Concepts are nothing but the locations in conceptual space and creativity is the act of identifying new locations within that space. Since kāyadā compositions are based on some specific rule structure, the new composition is said to be creative within the given structure.

Finally, transformational creativity which involves changing the rules which delimit the conceptual space, in turn it identifies the new sub-space. The proposed system gives birth to previously unexpected / impossible compositions which are of much greater surprise. The extreme transformations which do not have any relationship between the original rules set are eliminated with the help of fitness function.

**Pearson correlation coefficient**

The experiments with generated new compositions are carried out to see the correlation between the fitness value given by the machine and fitness value given by human music expert. In this experimental setup, exhaustive tests were carried out by considering top 20 compositions, which are given high fitness by machine, of different 5 themes of kāyadā. These 20 compositions are given to music expert to evaluate and to give score according to their musical genre and quality.

We use Pearson correlation coefficient to show the linear correlation between the fitness given by our system and human music expert. Equation 3 defines the sample correlation coefficient formula.

$$r = \frac{\sum_{i=1}^{n}(x_i - \bar{x})(y_i - \bar{y})}{\sqrt{\sum_{i=1}^{n}(x_i - \bar{x})^2} \sqrt{\sum_{i=1}^{n}(y_i - \bar{y})^2}} \qquad (3)$$

Here, '$r$' is Pearson correlation coefficient between two variables, 'n' is the sample size, $x_i$ and $y_i$ are the individual samples from paired data (X, Y), X: quality of composition rated by music expert and Y: quality of composition given by machine.

Table 8 shows the results obtained of this experiment. The value of '*r*' which is close to 1, implies a linear relationship between X and Y.

**Table. 8.** Theme composition and respective Pearson correlation coefficient

| THEME # | R |
|---|---|
| 1 | 0.77 |
| 2 | 0.85 |
| 3 | 0.93 |
| 4 | 0.92 |
| 5 | 0.88 |

**Fitness function as creativity measure**

In MA the initial population of composition is generated randomly, therefore there was lesser chance of having the compositions with good quality, and our objective was to generate as many valid compositions which are good in quality and which are novel with fitness zero (minimization problem). We wanted to see the effectiveness of proposed methodology across different generation numbers and with different population sizes. For that we have taken the generation number along the X-axis and percentage of compositions with the fitness zero along Y-axis. Therefore, in each generation we have taken the proportionate of compositions out of the total compositions having fitness value zero.

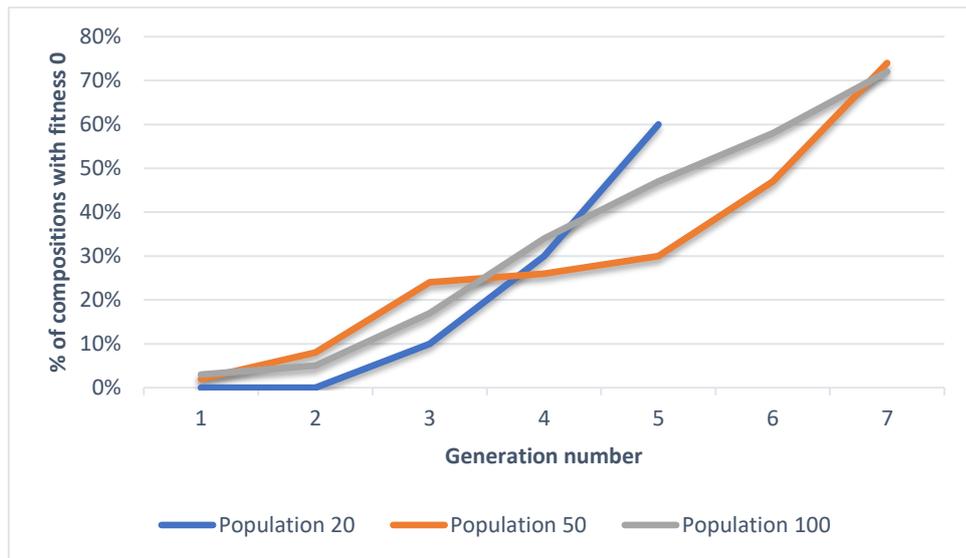

**Fig. 21.** Fitness value for various population across different generations

We have seen that initially the value comes out to be zero or very minimal, but as generation number increases it increases exponentially. This shows that the methodology is helpful in producing items dissimilar to existing examples of that genre. From the graph one can see that when we stop the process after seventh generations, we could be able to have almost 60%-70% compositions generated are of good quality and novel compositions with fitness value zero.

**Test of independence**

We wanted to investigate whether there is a statistically significant difference between compositions which are generated by machine and compositions which are generated by actual experts in the music domain. The level of significance of difference between compositions generated by proposed system and compositions generated by human expert is statistically determined by a chi-square test.

For that we have taken 20 different compositions out of which 10 compositions were valid compositions taken from experts and 10 were generated by machine (compositions which were having high fitness values).

We have conducted a survey among 10 different experts who had proficiency in the field of music for around 10-15 years and had graduated with the degree equivalent to bachelor's degree in the field of tabla. It was not revealed to them that which compositions were computer generated and which were human generated. We played the compositions one by one and asked the experts to identify whether it was a human generated one or machine generated one.

To analyze how well the observed distribution of data fit with the distribution that is expected, we carried out chi-square test of independence. Firstly, we calculated the expected value of two variables namely, human generated compositions (HGC) and computer-generated compositions (CGC). The table 9 shows the observed values of these two variables by different music experts:

**Table. 9.** Observed Frequencies

|  | Expert1 | Expert2 | Expert3 | Expert4 | Expert5 | Expert6 | Expert7 | Expert8 | Expert9 | Expert10 | Total |
|---|---|---|---|---|---|---|---|---|---|---|---|
| **HGC** | 13 | 12 | 11 | 11 | 18 | 11 | 13 | 14 | 15 | 16 | **134** |
| **CGC** | 7 | 8 | 9 | 9 | 2 | 9 | 7 | 6 | 5 | 4 | **66** |
|  |  |  |  |  |  |  |  |  |  |  |  |
| **Total** | 20 | 20 | 20 | 20 | 20 | 20 | 20 | 20 | 20 | 20 | **200** |

Applying the chi-square test for independence to sample data, we compute the degrees of freedom, the expected frequency counts, and the chi-square test statistic.

We calculate the expected values for these two variables using formula:

$$E_{i,j} = \frac{N_i N_j}{N} \tag{4}$$

where, $N_i$ = Sum of i<sup>th</sup> column

$N_j$ = Sum of j<sup>th</sup> column

N = Total Number of observations in the sample

**Table. 10.** Expected frequencies

|  | Expert1 | Expert2 | Expert3 | Expert4 | Expert5 | Expert6 | Expert7 | Expert8 | Expert9 | Expert10 |
|---|---|---|---|---|---|---|---|---|---|---|
| **HGC** | 13.4 | 13.4 | 13.4 | 13.4 | 13.4 | 13.4 | 13.4 | 13.4 | 13.4 | 13.4 |
| **CGC** | 6.6 | 6.6 | 6.6 | 6.6 | 6.6 | 6.6 | 6.6 | 6.6 | 6.6 | 6.6 |

Degree of freedom is calculated by using formula:

DF = (r - 1) * (c - 1) = (2 - 1) * (10 - 1) = 1 * 9 = 9

where, DF = Degree of freedom

r = number of rows

c = number of columns

We state the null hypothesis as follows:

H₀: There is no association between of expert identification of compositions and actual nature of data.

And alternative hypothesis is as,

H₁: There is significant association between these two.

$$X^2 = \sum_{i=1}^{r} \sum_{j=1}^{c} \frac{(O_{i,j} - E_{i,j})^2}{E_{i,j}} \quad (5)$$

Where, $X^2$ = Chi-square test of independence

$O_{i,j}$ = Observed values of two variables

$E_{i,j}$ = Expected values of two variables

After calculation we get the value of $X^2$ as 11.39756

We calculated the P value, which returns the probability that a value of the $X^2$ statistic at least as high as the value calculated by the above formula could have happened by chance under the assumption of independence.

We use the chi-square test to find P ($X^2$ > 11.39) = 0.90974

Based on the result of chi-square test, we conclude:

Since the calculated P-value (0.90974) is more than the significance level (0.05), we accept null hypothesis H₀ and conclude that there is no association between two variables namely HGC and CGC.

**Evaluation by music experts**

In order to validate our results, we further conducted a survey among the ten experts. We played the compositions which are generated by machine one by one and asked the experts to give score between 1-10 depending upon novelty and quality of the composition.

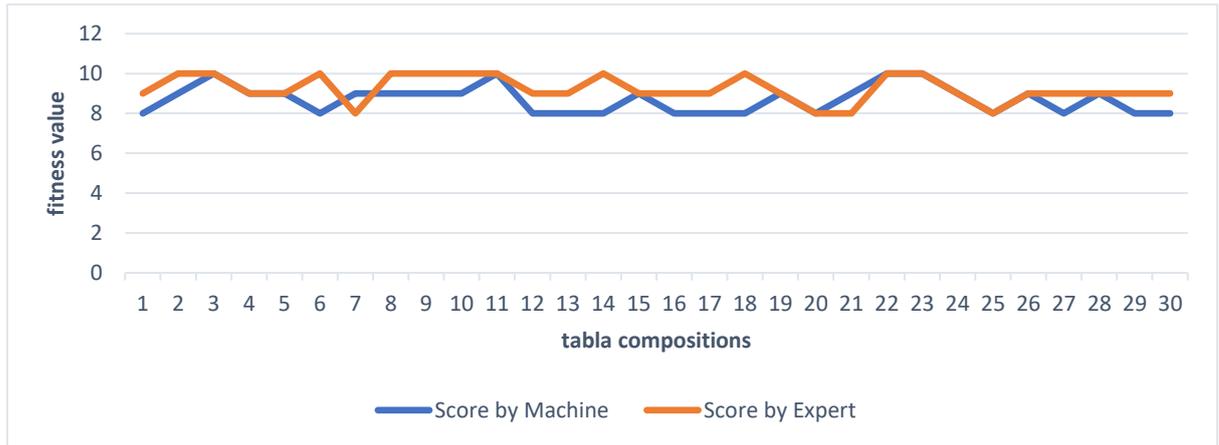

**Fig. 22.** Fitness values given by machine vs by music expert

A higher value indicates the good quality and novel composition. Figure 22 shows the scores given by machine (normalized fitness values in a scale from 1 to 10) and scores by the music experts. We could see that human experts also evaluated the compositions almost in the similar fashion as recommended by the machine. It shows that the proposed methodology could come out with compositions, which are valid, legal, novel and of good quality.

**Evaluation of Ritchie's criteria**

On the basis of the results obtained from the creative system it was essential that we evaluated our system with some principles which are used to assess the computer program. Graeme Ritchie proposed a set of fourteen criteria to assess the creative systems. (Ritchie, 2007)

Ritchie's fourteen criteria for assessing creativity are:

1. Average typicality

2. Ratio typical results / all results

3. Average quality

4. Ratio good results / all results

5. Ratio good typical results / all results

6. Ratio good atypical results / all results

7. Ratio good atypical results / atypical results

8. Ratio good atypical results / good typical results

9. Ratio results in the inspiring set / inspiring set

10. Ratio all results / results in the inspiring set

11. Average typicality of new results

12. Average quality of new results

13. Typical new results / new results

14. Good new results / result

There are few parameters on which Ritchie's criteria is based on:

**Basic item B:** an entity that program produces. In the proposed system the basic items produced are the different variations of the given kāyadā theme composition.

**Inspiring set I:** the set of basic items that implicitly or explicitly drive the development of the program. In our example the theme of a kāyadā composition is nothing but the inspiring set.

**R**: The set of results produced by the system.

**typ:** typicality of the items. The generated variation is said to be typical if it has satisfied the rules of the composition and has syntactically correct formation.

**val:** value of the items. The generated variation is said to be good if the composition appeals to the music expert with the creative and innovative nature of it.

There are sets of parameters which are used as initialization data for the system:

- Kāyadā theme

- Rules Set

With every iteration of the methodology the system produces different variations of the kāyadā theme. The system stops when it meets the stopping criteria (maximum iterations or when it reaches the goal fitness) by producing results which are prominent.

In order to obtain the results which are later on analyzed by Ritchie's criteria, the set of five different themes of kāyadā composition are studied with the initialization data specified above. For each theme 20 different variations are considered which are above the fitness threshold. Music experts, who are trained as tabla players were asked to rate the generated variations on two scales: typicality and quality.

For the evaluation purpose we set the initial values of the parameters as:

1. Typicality threshold: 0.7

2. Quality threshold: 0.7

3. Total number of results obtained by the system: 20

4. Total number of items present in inspiring set: 1

The results obtained after applying Ritchie's criteria are shown in the table 11.

**Table. 11.** Criteria and their values with respect to various theme compositions

| CRITERIA NUMBER | | THEME 1 | THEME 2 | THEME 3 | THEME 4 | THEME 5 |
|---|---|---|---|---|---|---|
| 1. | Average typicality | 0.98 | 0.99 | 0.98 | 0.97 | 0.97 |
| 2. | Ratio typical results / all results | 1 | 1 | 1 | 1 | 1 |
| 3. | Average quality | 0.7 | 0.76 | 0.79 | 0.71 | 0.75 |
| 4. | Ratio good results / all results | 0.65 | 0.9 | 0.85 | 0.6 | 0.75 |
| 5. | Ratio good typical results / all results | 1 | 1 | 1 | 1 | 1 |
| 6. | Ratio good atypical results / all results | 0 | 0 | 0 | 0 | 0 |
| 7. | Ratio good atypical results / atypical results | ∞ | ∞ | ∞ | ∞ | ∞ |
| 8. | Ratio good atypical results / good typical results | 0 | 0 | 0 | 0 | 0 |
| 9. | Ratio results in the inspiring set / inspiring set | 0 | 0 | 0 | 0 | 0 |
| 10. | Ratio all results / results in the inspiring set | ∞ | ∞ | ∞ | ∞ | ∞ |
| 11. | Average typicality of new results | 0.98 | 0.99 | 0.98 | 0.97 | 0.97 |
| 12. | Average quality of new results | 0.7 | 0.76 | 0.79 | 0.71 | 0.75 |
| 13. | Typical new results / new results | 1 | 1 | 1 | 1 | 1 |
| 14. | Good new results / result | 0.65 | 0.9 | 0.85 | 0.6 | 0.75 |

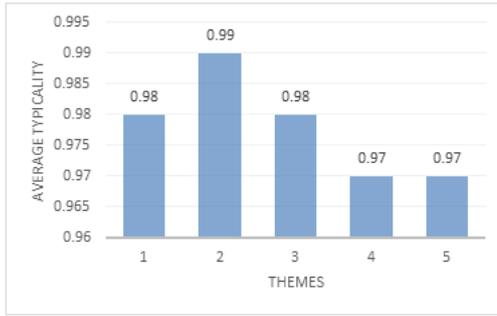 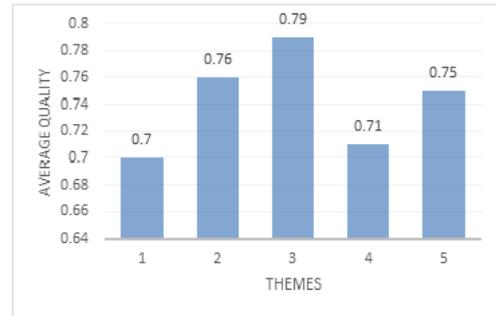

**Fig. 23.** Average typicality of 5 compositions  **Fig. 24.** Average quality of 5 compositions

After analyzing the table 11 we can notice that, the proposed systems' results with typicality and quality (criteria 1 and 3) of compositions are above 97% and 70% respectively. This demonstrates that the system could be able to generate novel and good quality compositions.

There were other works related to creativity apart from music domain that applied Ritchie's criteria to assess the creativity of the system. We compared our computational creative model with them and results are described in table 12.

**Table. 12.** Comparison with other creative systems

| RITCHIE'S CRITERIA | WASP | DIVAGO | DUPOND | CBVD | PROPOSED SYSTEM |
|---|---|---|---|---|---|
| 1. Average typicality | 0.71 | 0.54 | 0.56 | 0.74 | 0.98 |
| 2. Ratio typical results / all results | 0.54 | 0.56 | 0.75 | 0.8 | 1 |
| 3. Average quality | 0.47 | 0.78 | 0.29 | N/A | 0.7 |
| 4. Ratio good results / all results | 0.24 | 0.78 | 0.1 | N/A | 0.65 |
| 5. Ratio good typical results / all results | 0.36 | 0.78 | 0.1 | N/A | 1 |
| 6. Ratio good atypical results / all results | 0.05 | 0.34 | 0.02 | N/A | 0 |
| 7. Ratio good atypical results / atypical results | 0.12 | 0.79 | 0.09 | N/A | ∞ |
| 8. Ratio good atypical results / good typical results | 0.28 | 0.79 | 0.33 | N/A | 0 |
| 9. Ratio results in the inspiring set / inspiring set | 0 | 0.03 | 0.01 | 0 | 0 |
| 10. Ratio all results / results in the inspiring set | ∞ | 0.93 | 0.98 | 0 | ∞ |
| 11. Average typicality of new results | 0.71 | 0.51 | 0.56 | 0.74 | 0.98 |
| 12. Average quality of new results | 0.47 | 0.83 | 0.29 | N/A | 0.7 |
| 13. Typical new results / new results | 0.54 | 0.5 | 0.75 | 0.8 | 1 |
| 14. Good new results / result | 0.24 | 0.78 | 0.1 | N/A | 0.65 |

1. WASP: The WASP system draws on prior poems and a selection of vocabulary provided by the user to generate a metrically driven re-combination of the given vocabulary according to the line patterns extracted from prior poems (Gervas, 2000).

2. Divago: Divago is a system for Concept Invention that aims to generate concepts via a mechanism of Conceptual Blending (Pereira, 2005).

3. Dupond: Once given a sentence and a set of configuration options to Dupond system, it parses that sentence, disambiguates the words and replaces some of them by synonyms or hypernyms. The output sentences are different from the input ones, keeping the original meaning unchanged (Mendes & Pereira, 2004).

4. CBVD: Conceptual blending for the visual domain (CBVD) is a framework that formalizes the entire process of conceptual blending while applying it to the visual domain (Steinbrück, 2013).

Different creative domains such as linguistic creativity, artistic creativity and musical creativity have been considered for the comparison. The detailed evaluation and comparison in table 12 depicts that the proposed system individually and with comparison with others stand out and demonstrate perceived creativity. Specially the criteria's like typicality and quality achieve high results in comparison with other systems.

**Pease et al.'s tests on the input, output and process of a system**

Pease et al. (2001) proposed a combination of tests which are used to evaluate the creativity of artificial systems. Factors like input provided to the system, output generated from the system and the overall process employed by the system are considered for the evaluation of the creative system.

For the proposed system following aspects are considered:

1. Input given to the system: a system is said to be creative if it produces items which are not part of the inspiring set (input set).

2. Output produced by the system: the output of the system is evaluated based on fundamental novelty, complexity and novelty of the creative process, typicality, surprising, novelty and quality perception by human expert.

3. Process: the randomness is measured for the creative process using probabilities of items being generated. Also, it should measure the correlation of self-evaluation and external evaluation of generated output. Some quality measures should be compared between two sets of output, where the quality of first should exceed the other.

We observe that the proposed system evaluates and reflects above mentioned aspects of creativity of artificial systems. The first criteria is evaluated empirically with an experiment with the human expert, where experts mentioned that, final improvised compositions which are generated after successful completion of a process were not part of the input set. The proposed system qualifies all aspects based on novelty of the output composition, typicality and quality of the generated compositions. Criteria three demonstrates the effectiveness of fitness function to improve compositions quality across different generations.

**Colton's creative tripod framework**

The creative tripod (Colton, 2008) represents three qualities that a creative system must demonstrate to some degree:

1. Skill

2. Imagination

3. Appreciation

If a creative system can demonstrate each of these three behaviors, then Colton argues that this is sufficient for the system to be perceived as creative. Demonstration of above mentioned three behaviors with proposed methodology are explained as follows:

- Skill: The proposed methodology can generate compositions which are different from each other but belong to same genre.

- Imagination: The methodology takes an evolutionary approach to generation of different compositions and can generate compositions which are not seen before.

- Appreciation: The proposed methodology can recognize the good quality compositions using fitness function which makes the system autonomous in the creative behavior.

In order to maintain the balance of the tripod analogy, a system needs to be well-presented with all three qualities. The proposed system maintains the balance between all three and hence it is highly skillful with the all three qualities.

**IV Conclusion**

We have presented a genuinely creative system which can effectively create new types of variations by preserving semantic properties of the genre. The system provides an experimental setup that enables testing various hypothesis and evaluates the properties of creativity.

Our main contribution has been to introduce a methodology which acts as a creative framework for percussion music improviser. Rather than only assessing the creative product we propose to assess the impact of other key terms like producer, process and environment which contribute in the computational creativity.

The findings reinforce the emergence of machine-driven creativity demonstrated in the methodology. The results were also compared with those of other known computational creative systems available to show the supremacy of the methodology used.